\documentclass[a4paper,11pt]{article}

\usepackage{jcappub}
\usepackage[utf8]{inputenc}

\usepackage{siunitx}

\usepackage{graphicx}
\usepackage{xcolor}

\usepackage{caption}
\usepackage{subcaption}
\usepackage{amsmath}
\usepackage{amssymb}
\usepackage{float}
\usepackage{threeparttable}

\usepackage{natbib}
\def\be{\begin{equation}}
\def\ee{\end{equation}}
\def\bea{\begin{eqnarray}}
\def\eea{\end{eqnarray}}

\def\s{{\rm s}}
\def\yr{{\rm yr}}
\def\cm{{\rm cm}}

\def\mpc{{\rm Mpc}}
\def\gev{{\rm GeV}}

\title{
Searching for velocity-dependent dark matter annihilation signals from extragalactic halos
}
\author[a]{Eric J.~Baxter,}
\author[b]{Jason Kumar,}
\author[b]{Aleczander D. Paul,}
\author[b]{Jack Runburg}

\affiliation[a]{\mbox{Institute for Astronomy, University of Hawai`i, 2680 Woodlawn Drive, Honolulu, HI 96822, USA}}
\affiliation[b]{\mbox{Department of Physics \& Astronomy, University of Hawai`i, Honolulu, HI 96822, USA}}

\emailAdd{runburg@hawaii.edu}

\abstract{We consider gamma-ray signals of dark matter annihilation in extragalactic halos in the case where dark matter annihilates from a $p$-wave or $d$-wave state.  In these scenarios, signals from extragalactic halos are enhanced relative to other targets, such as the Galactic Center or dwarf spheroidal galaxies, because the typical relative speed of the dark matter is larger in extragalactic halos.  We perform a mock data analysis of gamma rays produced by dark matter annihilation in halos detected by the Sloan Digital Sky Survey.  We include a model for uncorrelated galactic and extragalactic gamma ray backgrounds, as well as a simple model for backgrounds due to astrophysical processes in the extragalactic halos detected by the survey.  We find that, for models which are still allowed by other gamma ray searches, searches of extragalactic halos with the current Fermi exposure can produce evidence for dark matter annihilation, though it is difficult to distinguish the $p$-wave and $d$-wave scenarios.  With a factor $10\times$ 
larger exposure, though, discrimination of the velocity-dependence is possible.}

\begin{document}
\maketitle

\section{Introduction}

A promising strategy in dark matter indirect detection 
is the search for photons produced by dark matter annihilation in astrophysical objects with a large 
dark matter density.  This strategy is particularly fruitful 
because the photons point back to their source, allowing 
one to use prior information about the dark matter distribution to enhance detection prospects. 
For example, some of the best sensitivity to dark matter annihilation arises from searches for high energy photons arriving from 
the direction of dwarf spheroidal galaxies (dSphs)~\cite{Fermi-LAT:2010cni,Fermi-LAT:2011vow,Fermi-LAT:2013sme,Fermi-LAT:2015att} or 
from the Milky Way Galactic Center (GC)~\cite{Goodenough:2009gk,Hooper:2010mq,Abazajian:2012pn,Daylan:2014rsa,Fermi-LAT:2017opo}, which are localized on the sky, are thought to have large dark matter densities, and are relatively nearby.  

Extragalactic dark matter halos also contain large amounts of dark matter, and can be localized with observations of the galaxies that they host, gravitational lensing, and other techniques.  Relative to dSphs, the larger masses of extragalactic halos and their larger distances push the annihilation signal in opposite directions, leading to a total expected flux from known extragalactic halos that is similar to that from known dSphs in standard dark matter annihilation models.\footnote{To arrive at this claim, we assume that all extragalactic halos with mass $M \gtrsim 10^{13}\,M_{\odot}$ out to redshifts of $z \sim 1.0$ can be detected, and adopt the halo mass function from Ref.~\cite{Tinker:2008}.  As we discuss in more detail in \S\ref{sec:halo_mass_dependence}, if the velocity-weighted annihilation cross-section is velocity-independent, the annihilation luminosity is thought to scale roughly as $M^{0.86}$, allowing the extragalactic flux to be computed up to some overall normalization.  We compute the dSph flux up to the same overall normalization factor using rough estimates for the masses and distances of local dSphs.  We ignore a possible boost factor, which could significantly enhance the extragalactic contribution (see, for example, Ref.~\cite{Regis:2015zka}). }  To date, however, extragalactic dark matter halos have not provided constraints on the dark matter annihilation cross section that are competitive with those from dSphs.
There are two primary reasons for this.  First, dSphs take up a smaller sky fraction than extragalactic halos, meaning that the total background contribution for dSphs is significantly smaller.  Second, dSphs are expected to have smaller correlated astrophysical backgrounds since there are few conventional gamma-ray producing sources inside of these dark matter-dominated objects.  Extragalactic halos, on the other hand, can be powerful emitters of gamma-rays via non-dark matter processes \cite[e.g. Ref.][]{Stecker:2011}, making separating signal from background challenging (see further discussion of observational challenges of detecting annihilation radiation from extragalactic halos in Refs.~\cite{Sanchez-Conde:2011,Nezri:2012,Jeltema:2009}).

However, in dark matter models where the velocity-weighted annihilation cross section is velocity-dependent, the signal from extragalactic halos can be greatly enhanced relative to that from dSphs.  
Our particular focus is on scenarios in which the dark 
matter annihilation only occurs from a $p$-wave or 
$d$-wave initial state, scenarios realized in a variety 
of well-motivated theoretical models.  If dark matter 
annihilation is $p$-wave or $d$-wave suppressed, then 
the annihilation cross section is suppressed 
by factors 
of $(v/c)^2$ or $(v/c)^4$, respectively, where $v$ is the relative 
velocity of the dark matter particles.  Searches of large extragalactic halos are particularly 
interesting in this case, because the relative velocities 
of dark matter particles in these halos tend to be larger than in targets within the Milky Way, implying 
that the signal from extragalactic halos is enhanced 
relative to that from dSphs or the GC for $p$-/$d$-wave suppressed dark matter annihilation.   Previous studies of velocity-dependent dark matter annihilation in extragalactic dark matter halos include Refs.~\cite{Piccirillo:2022,Lacroix:2022}.

The scenario of $p$-wave or $d$-wave annihilation is especially 
interesting in the context of the Galactic Center (GC) 
excess~\cite{Goodenough:2009gk,Hooper:2010mq}.  This 
excess of GeV-range photons has been studied in the context 
of $s$-wave dark matter annihilation, where it is found that 
models which would be consistent with the GC excess are 
roughly at the limit of current bounds from searches for 
gamma rays from dwarf spheroidal galaxies (see, for example Refs.~\cite{Chang:2018bpt,Hooper:2019xss}).  
But in models 
of $p$- or $d$-wave annihilation, the constraints from 
dSphs are weakened relative to those from the GC, because 
relative particle speeds tend to be smaller in dSphs, yielding 
reduced annihilation rates~\cite{Boddy:2018ike}.  As a result, the scenario of 
$p$- or $d$-wave annihilation would weaken 
constraints from dSphs on dark matter explanations of the GC excess.  But since these scenarios yield enhanced annihilation 
rates in extragalactic halos, those become an important search 
target for exploring explanations of the GC excess.

In this paper, we perform an idealized mock analysis to explore the prospects for detecting and characterizing the annihilation signal from
extragalactic dark matter halos  
in models where the annihilation cross section is $p$-wave or $d$-wave suppressed.  We generate mock gamma-ray data using a catalog of real halos detected by the Sloan Digital Sky Survey, and analyze this data  using a template-based approach.  
While our approach is highly idealized compared to an analysis of real data, it suffices to determine the information content that could conceivably be extracted about $p$- and $d$-wave dark matter annihilation from current and future gamma-ray data sets.
We will find that the simple intuition from above holds: in the idealized scenarios considered here, analyzing the extragalactic dark matter signal yields significantly tighter constraints on these models than analyses of dSphs.  We find that this result is at least somewhat robust to freedom in background models.

The plan of this paper is as follows.  In \S\ref{sec:halocatalog} we describe the catalog of extragalactic halos we use in our analysis; in \S\ref{sec:model} we describe our model for the dark matter annihilation signal; we describe the creation of a mock data set in \S\ref{sec:mock_data}, and the results of analyzing this data are described in \S\ref{sec:results}.  We conclude in \S\ref{sec:conclusion}.

\section{Catalog of Extragalactic Dark Matter Halos}
\label{sec:halocatalog}

For the purposes of generating mock data, we rely on a catalog of dark matter halos detected in real data.  Massive dark matter halos host galaxies, making galaxy surveys a powerful tool for identifying and localizing halos.  By identifying galaxies that live in close proximity to one another via a spectroscopic survey, one can assign galaxies to groups, i.e. collections of galaxies that live in the same dark matter halo.  
In order to estimate the amplitude of the annihilation signal from each halo, we must have some knowledge of the halo's mass distribution and distance. To this end, halo masses can be estimated from a variety of techniques, such as weak lensing or clustering.  The distance, on the other hand, can be computed from the halo's redshift assuming a cosmological model.

In this analysis, we use the galaxy group/halo catalog presented in Ref.~\cite{Lim:2017}, which is based on data from the Sloan Digital Sky Survey (SDSS).\footnote{\url{https://www.sdss.org/}}  This halo catalog uses a modified version of the algorithm introduced in Refs.~\cite{Yang:2005, Yang:2007} to assign galaxies to groups and estimate halo masses.  We describe the method here briefly and refer readers to Ref.~\cite{Lim:2017} for more details.  First, galaxies are assigned a preliminary group identification assuming that every galaxy lives in its own halo, with halo mass computed based on observational proxies including stellar mass and luminosity.  Next, assuming the group is described by a Navarro-Frenk-White (NFW) profile \cite{NFW}, the estimated mass is then used to estimate a halo radius and velocity dispersion, and the group membership is updated.  Halo centers are then recomputed using the updated group membership.  Next, halo mass estimates are recomputed using abundance matching, and the process is repeated.  Ref.~\cite{Lim:2017} assumes the $\Lambda$CDM cosmological model favored by WMAP9 \cite{Bennett:2013} when constructing their catalog. 

\begin{figure}[H]
    \centering
    \includegraphics[width=0.7\textwidth]{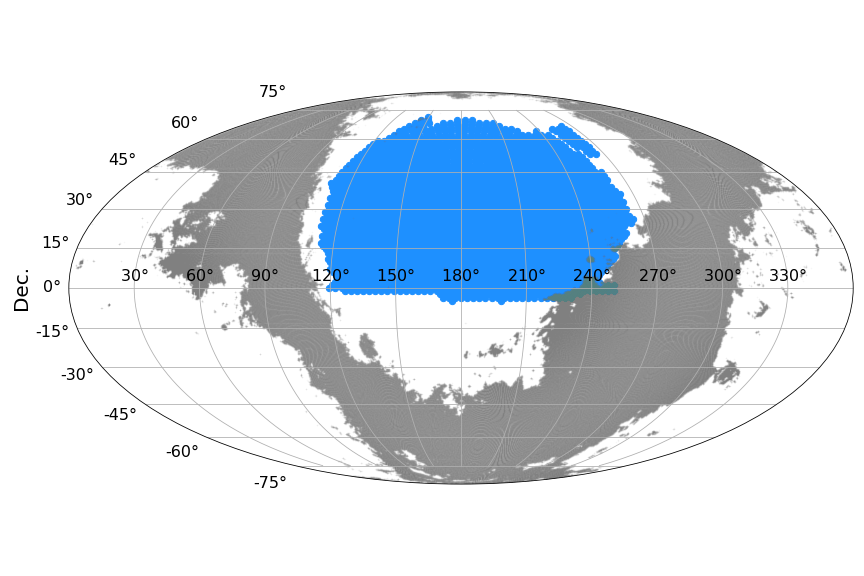}
    \caption{Sky footprint in equatorial coordinates of the halo catalog used in our analysis (blue).  This catalog is from Ref.~\cite{Lim:2017}, and derived from SDSS data.  Also shown in grey is the region of sky that is expected to have a galactic gamma-ray flux above the 75th percentile, as estimated using the model described in \S\ref{sec:gal_bg}.
    }
    \label{fig:survey_geometry_skymap}
\end{figure}

In order to remove fluctuations in the halo abundance that are due to variations in survey depth, and to make our results less dependent on survey strategy, we impose a completeness requirement on the halo catalog.  Ref.~\cite{Lim:2017} find that their SDSS-based halo catalog is complete for halo masses above $M \gtrsim 10^{12.5}\,M_{\odot}/h$ out to $z \lesssim 0.13$, and we select halos that satisfy these conditions.  The range of redshifts in the resulting catalog is then $0.006 \lesssim z \lesssim 0.13$, with median $\tilde{z} = 0.1$.  The range of halo masses in the catalog is $5 \times 10^{12}  \lesssim M/M_{\odot} \lesssim 2\times 10^{15} $, with median $\tilde{M} = 5 \times 10^{12} M_{\odot}$.  There are a total of 59198 halos in the catalog.  The sky footprint of the resulting catalog is presented in Figure~\ref{fig:survey_geometry_skymap}.

Our mock data set will miss contributions to the annihilation radiation that come from halos outside the mass and redshift bounds that we impose.  Halos at higher redshifts will be largely uncorrelated with the halos in our catalog, and so are expected to act as a noise source for our analysis.  We discuss our noise and background models in \S\ref{sec:model}.   Halos in the redshift range that we consider, but less massive than our mass cut, will produce a background of annihilation radiation that is correlated with the signal from high-mass halos.  However, for the dark matter models that we consider, the fraction of annihilation radiation that we exclude is expected to be small.  For the $n=2$ model considered below, our mass cut removes only about 10\% of the expected annihilation luminosity at $z=0$, assuming the halo mass function from Ref.~\cite{Tinker:2008}; for the $n= 4$ model, the mass cut removes less than 1\% of the annihilation luminosity at $z=0$ (see further discussion of these points in \S\ref{sec:halo_mass_dependence}).

\section{Signals and Backgrounds}
\label{sec:model}

\subsection{Expected annihilation flux from a dark matter halo}

The flux of photons arising from dark matter annihilation in any dark matter halo can 
be expressed as 
\bea
\Phi_{\rm DM} &=& \Phi_{PP} \times J ,
\eea
where 
$\Phi_{PP}$ depends only on information about the dark matter particle 
physics, while $J$ (the $J$-factor) depends on the astrophysics 
of the target, and the velocity-dependence of the cross section. 
 In particular, 
\bea
\Phi_{PP} &=& \frac{(\sigma v)_0}{8\pi m_X^2} 
\times \int_{E_{\rm min}}^{E_{\rm max}} dE_\gamma~\frac{dN_\gamma}{dE_\gamma} ,
\nonumber\\
J &=& \frac{1}{D^2} 
\int d^3 v_1 \int d^3 v_2 ~ 
f(\vec{r}_1, \vec{v}_1) f(\vec{r}_2, \vec{v}_2) 
\times S(|\vec{v}_1 - \vec{v}_2|/c) ,
\label{eqn:J}
\eea
where $m_X$ is the dark matter mass, 
$\sigma v = (\sigma v)_0 \times S(v/c)$ 
is the dark matter 
annihilation cross section, $v$ is the relative velocity 
between dark matter particles, and $dN_\gamma / dE_\gamma$ is the energy spectrum of photons per annihilation
(see, for example,~\cite{Boddy:2017vpe}).
The limits of the integral over $E_{\gamma}$ are restricted to $E_{\rm min}$ to $E_{\rm max}$, corresponding to the energy range of the detector.
We assume that the dark matter particle is its own 
anti-particle.
$D$ is the distance to the halo, and $f(\vec{r}, \vec{v})$ is the dark matter velocity distribution function.  The dark 
matter density distribution is then given by 
$\rho (\vec{r}) = \int d^3 v~f(\vec{r}, \vec{v})$.
Note, in writing Eq.~\ref{eqn:J}, we have assumed that the 
size of the halo is much smaller than the distance $D$ to the 
halo.  This is an excellent approximation for the extragalactic 
halos that we will consider.

We will assume that all extragalactic dark matter halos have velocity distributions 
which are spherically-symmetric and isotropic, and depend on only two parameters: 
a scale density $\rho_s$ and a scale radius $r_s$.  We thus assume that each halo has 
a velocity-distribution which depends only on $r = |\vec{r}|$ and $v = |\vec{v}|$, and 
the velocity-distributions of two halos differ only by the values of the parameters 
$\rho_s$ and $r_s$.  In reality, dark matter halos may have more complicated density and velocity distributions \cite[e.g.][]{Jing:2002}, including the effects of triaxiality, velocity anisotropies, and detailed effects from 
baryonic physics.  But we will see that these effects will not alter our overall conclusions, because 
the effect of their variation from halo to halo will be relatively small compared to that of 
the variation of the 
parameters $\rho_s$ and $r_s$.  We note also that the supermassive black holes living at the centers of galaxies could result in changes to the dark matter distribution, and the formation of a dark matter `spike' \cite{Ullio:2001}; the impact of the velocity-dependent annihilation on such spikes has been considered in Refs.~\cite{Amin:2008,Sandick:2016zeg}.

We are interested in the case in which  the velocity-dependence of the dark matter annihilation 
cross section is power-law ($S(v/c) = (v/c)^n$).  
The most interesting cases, for our purpose, are 
$p$-wave ($n=2$) and $d$-wave ($n=4$) annihilation, 
since the annihilation rate grows with relative velocity.
Both of these examples are theoretically well-motivated.  
For example, 
dark matter annihilation will be $p$-wave if dark matter is a 
fermion which annihilates through an $s$-channel scalar 
interaction.  
$p$-wave annihilation can also be the 
dominant process if dark matter is a Majorana fermion 
which annihilates to a Standard Model fermion/anti-fermion pair through any interaction which respects minimal flavor violation (MFV) (see, for example, Ref.~\cite{Kumar:2013iva}).
Similarly, $d$-wave annihilation may be 
the dominant process if dark matter is instead a real scalar 
particle~\cite{Kumar:2013iva,Giacchino:2013bta,Toma:2013bka}. Models with $n=1$ could arise in so-called `Impeded Dark Matter' scenarios \cite{Kopp:2016}, but for simplicity, 
we will not consider that scenario here.

Under these assumptions, all of the dependence of the 
$J$-factor on the parameters can be determined by 
dimensional analysis~\cite{Boddy:2019wfg}.  In particular we find
\bea
J &=& 
\frac{4\pi M_s^2 }{D^2 r_s^3} 
\left(\frac{4\pi G_N M_s  / r_s}{c^2} \right)^{n/2} 
\times \tilde J_n^{\rm tot} ,
\nonumber\\
&=& 
\frac{1.8~\cm^{-2} \yr^{-1}}{(\gev /c^{-2})^{-2} \cm^3 \s^{-1}} 
\left[\frac{M_s}{M_\odot}\right]^2  
\left[\frac{r_s}{\mpc}\right]^{-3}
\left[\frac{D}{\mpc}\right]^{-2}
\left(6.1\times 10^{-19} 
\left[\frac{M_s}{M_\odot}\right]  
\left[\frac{r_s}{\mpc}\right]^{-1}
\right)^{n/2} 
\nonumber\\
&\,& 
\times \tilde J_n^{\rm tot}  ,
\nonumber \label{eq:J_factor} \\
\eea
where $M_s \equiv \rho_s r_s^3$, and  
$\tilde J_n^{\rm tot}$ is a constant. 
The dependence of $J$ on $r_s$ and $M_s$ in Eq.~\ref{eq:J_factor}, which determines our main results, will not depend on the functional form of the dark matter velocity distribution.  The form of the velocity distribution will 
affect $\tilde J_n^{\rm tot}$, and this quantity has been 
estimated using a variety of techniques (for previous 
work on determining the $J$-factors for velocity-dependent 
dark matter matter annihilation, see~\cite{PhysRevD.79.083525,Belotsky_2014,Ferrer_2013,Boddy:2017vpe,PhysRevD.97.063013,Peta__2018,Boddy:2018ike,Lacroix_2018,Boddy:2019wfg,Boddy:2019qak,Boucher:2021mii}).
But this is an overall normalization for the flux from all halos, 
which is degenerate with $\Phi_{PP}$.  The relative magnitude of the fluxes from different 
halos depends only on the halo parameters $\rho_s$ and $r_s$, independent of the functional form of the halo density profile.  

To determine $J$ then requires constraints on $D$, $M_s$, and $r_s$ for each halo.  As mentioned previously, $D$ can be determined from the halo redshift assuming a cosmological model.  However, the catalog of Ref.~\cite{Lim:2017} does not provide independent measures of $M_s$ and $r_s$.  Rather, total halo masses are inferred from an iterative process that assumes the distribution of galaxies in each halo follows an NFW profile 
with a fixed relation between $\rho_s$ and $r_s$.
We will therefore adopt similar assumptions in our analysis.

The NFW density profile is given by 
\bea
\rho (r) &=& \frac{\rho_s}{(r/r_s) (1+ (r/r_s))^2} ,
\eea
where $r_s$ and $\rho_s$ are parameters specific to each halo, as discussed above.  The halo catalog of Ref.~\cite{Lim:2017} adopts the spherical overdensity convention for defining halo mass.  With this convention, the halo radius, $r_{\Delta,m/c}$, is defined such that the mass, $M_{\Delta, m/c}$, enclosed within $r_{\Delta,m/c}$ is a fixed multiple $\Delta$ of the (c)ritical or (m)ean density, $\rho_{m/c}$, of the Universe at the halo's redshift, $z$:
\begin{equation}
    M_{\rm halo} \equiv M_{\Delta,m/c} = \Delta \frac{4}{3}\pi r_{\Delta,m/c}^3 \rho_{m/c}(z).
\end{equation}
The halo catalog Ref.~\cite{Lim:2017} uses $\Delta = 180$ and $\rho_m$.  The halo mass can be related to $M_s$ using the fact that
\bea
M_{\Delta,m/c} &=& 4\pi \int_0^{r_{\Delta,m/c}} dr~ r^2~\rho(r) ,
\nonumber\\
&=& 4\pi M_s \left[\ln (1+c_{\Delta, m/c})  + 
\frac{1}{1+c_{\Delta, m/c}} -1 \right],
\eea
where we have defined the halo concentration parameter, $c_{\Delta, m/c}$, via $c_{\Delta, m/c} \equiv r_{\Delta, m/c} /r_s$.  The mean relationship between $M_{\Delta,m/c}$ and $c_{\Delta, m/c}$ for NFW halos can be calibrated from CDM simulations.  Here we use the relationship measured by Ref.~\cite{Duffy:2008} assuming $\Delta = 200$:
\begin{equation}
c_{200, c} = A(M_{200,c}/M_{\rm pivot})^B (1+z)^C
\end{equation}
where $M_{\rm pivot} = 2\times 10^{12}\,h^{-1}M_{\odot}$, $A = 5.71$, $B = -0.084$, $C = -0.47$; using an alternative concentration model, such as that from \cite{Diemer:2015}, results in a negligible change to our results.  
We now have a way to relate the halo masses reported by Ref.~\cite{Lim:2017} to $M_s$ and $r_s$, thereby enabling computation of the $J$-factors, and thus the expected number of photons due to dark matter annihilation from each halo.  We use the code \texttt{Colossus} \cite{COLOSSUS} to convert between different halo mass definitions.  Ignoring the distance dependence for simplicity, these relationships yield at $z=0$
\begin{eqnarray}
    J &\propto& \left[ \frac{M_s}{M_{\odot}} \right]^{2+n/2}
\left[\frac{r_s}{\mpc}\right]^{-3-n/2} \\
&\approx& \left[0.55\times 10^{12}\,M_{\odot}\left(\frac{M_{180m}}{10^{13} \,M_\odot} \right)^{1.05} \right]^{2+n/2} \left[0.093 \,{\rm Mpc}\left( \frac{M_{180m}}{10^{13} \,M_{\odot}} \right)^{0.415}\right]^{-3-n/2} \\
&\propto& M_{180m}^{0.86+0.32n} \label{eq:simplified_J}.
\end{eqnarray}

The above expression ignores scatter in the relationship between halo mass and concentration, which is known to exist as a result of the different assembly histories of halos \cite{Zhao:2003}.  
Such scatter will in turn introduce scatter into the annihilation luminosities of the halos.   However, this scatter is expected to be small.  For example, Ref.~\cite{Diemer:2015} find that the scatter in the mass-concentration relation is roughly 0.16 dex, independent of halo mass, redshift and mass definition.  This level of scatter is much smaller than than the variation in halo concentration across the range of halo masses that we consider.  Similarly,
in Ref.~\cite{zahid2018a}, it was found that the dark matter 
velocity dispersion has a power-law relationship with the 
halo mass, with very little scatter, for central halos found in 
Illustris simulations spanning more than three orders of 
magnitude in mass.  Since the velocity dispersion is 
proportional to $(G_N M_s / r_s)$, this corresponds to the 
statement that there is very little scatter in the 
$r_s - M_s$ relationship.

More generally, we have assumed that the functional form of the dark matter density profile depends only on the parameters 
$r_s$ and $M_s$. In order for either the mass-concentration 
relationship or the mass-velocity dispersion relationship to provide a 
power-law relationship between $M_s$ and $r_s$, one must necessarily introduce a new mass scale 
and a new length scale.  These are effectively provided by 
$M_{\rm pivot}$ and $(M_{\rm pivot}/ \rho_{m/c})^{1/3}$.  
In order for there 
to be scatter in either the mass-concentration relationship 
or the mass-velocity dispersion relationship, there 
would have to be another dimensionless parameter which could 
vary from halo to halo at fixed $z$.  This parameter could 
then appear implicitly in the dark matter density profile;  
a triaxiality parameter would be an example of such a parameter.  This would amount to a dependence of 
$\tilde J^{\rm tot}_n$ on some dimensionless parameter 
which varied 
from halo to halo.
But the small magnitude of scatter in 
these relationships suggest that any additional 
dimensionless parameter only has a small variation over 
the halo mass range which we consider.  Thus, we may 
treat $\tilde J^{\rm tot}_n$ as essentially constant, despite 
possible deviations in the halo velocity distribution from 
the form we have considered.

Finally, the number of photons expected due to dark matter annihilation in a halo characterized by 
the parameters $M_s$, $r_s$ and $D$ is 
\bea
N_{\rm DM}^{({\rm halo})} &=&  \Phi_{\rm DM} \times ({\rm exposure}) ,
\nonumber\\
&=& 
1.8  \left[\frac{M_s}{M_\odot}\right]^2  
\left[\frac{r_s}{\mpc}\right]^{-3}
\left[\frac{D}{\mpc}\right]^{-2}
\left(6.1\times 10^{-19} 
\left[\frac{M_s}{M_\odot}\right]  
\left[\frac{r_s}{\mpc}\right]^{-1}
\right)^{n/2} 
\nonumber\\
&\,& 
\times \tilde J_n^{\rm tot} 
\times \frac{(\rm exposure)}{\cm^{2} \yr} \times 
\frac{\Phi_{PP}}{(\gev/c^{2})^{-2} \cm^3 \s^{-1}},
\eea
where, for halos described by NFW profiles (using Eddington inversion), one finds~\cite{Boddy:2019wfg,Boucher:2021mii}
\bea
\tilde J_{n=0}^{\rm tot} = 0.33, \qquad
\tilde J_{n=2}^{\rm tot} = 0.14, \qquad \tilde J_{n=4}^{\rm tot} = 0.12 .
\eea
Note that other well-motivated profiles are 
possible  (for reviews, see e.g. Refs.~\cite{Diemand:2011,Salucci:2019}), and would yield different values of 
$\tilde J_n^{\rm tot}$.  But 
the assumption of an NFW profile, and the choice 
of $\tilde J^{\rm tot}_n$, only 
affect the overall normalization of the flux, not 
the parametric dependence.

\subsection{Boost factors}
    
For this analysis, we assume that there is no boost factor enhancement of the photon flux due to substructure in the 
extragalactic halos.  Estimates for the boost factor associated with extragalactic halos vary widely, from ${\cal O}(1)$ to 
a few orders of magnitude~\cite{Sanchez-Conde:2013yxa,Regis:2015zka}.  But for our purposes, the important question is how the boost factor varies with halo mass.  If the boost factor is constant, then it 
only yields an overall rescaling of the fluxes from 
all halos, which is degenerate with $\Phi_{PP}$.

If the extragalactic halo is really only characterized by 
two parameters, $\rho_s$ and $r_s$, then one would expect 
the boost factor to be essentially independent of halo 
mass, since there is no other dimensionless quantity 
which depends only on $G_N$, $\rho_s$ 
and $r_s$.  This is consistent with results of Ref.~\cite{Sanchez-Conde:2013yxa}, for example, in which 
it was found that, for a variety of models, the boost factor 
varies by about an order of magnitude as the halo mass 
varies over nine orders of magnitude.

We thus expect that our assumption of negligible boost factor 
will not have a major effect on the main results of our analysis.  
Note that, for $n>0$, we expect the effect of the boost 
factor to be even smaller because particles bound 
to subhalos tend to have smaller relative velocities (see, e.g., Ref.~\cite{Piccirillo:2022}).
If $\sigma v$ grows with velocity, then the signal 
from subhalos will be suppressed relative to that from 
the central halo.\footnote{One could imagine a scenario for models with $n > 0$ where the presence of substructure actually suppresses the total halo annihilation luminosity because substructure shifts the relative velocity distribution of the dark matter particles to smaller values.  However, even in the extreme case that substructure produces no annihilation luminosity, we expect the suppression of the total halo luminosity to be small since the substructure mass fraction is known to be $f_{\rm sub} \lesssim 0.15$ for the relevant halo mass scales \cite{Giocoli:2010}.  }

\subsection{Extragalactic Anisotropic Background}
\label{sec:exgal_aniso}

Since galaxies can be powerful gamma-ray emitters, and since the halos in our sample host many galaxies, we expect these halos to emit gamma-rays not associated with dark matter annihilation.  In addition to galactic gamma-rays, processes such as accretion shocks around galaxy clusters could also produce  gamma-ray signals correlated with the halos in our sample \cite{Inoue:2005}.    
These sources of gamma-rays  constitute an important background for any search for gamma-ray photons produced by dark matter annihilation in extragalactic halos.  

We will assume that future multi-wavelength observations can be used to identify and remove bright sources of known astrophysical (i.e. non-dark matter) origin.  We additionally make the very simple assumption that the luminosity of 
the remaining astrophysical sources in an extragalactic halo is proportional 
to the halo mass.  Since we will be analyzing mock data generated 
from this model, it is not necessary that this model be precisely 
correct.  By assuming that the astrophysical gamma-ray emission is proportional to the halo mass, we
build in the fact that large halos, which are likely to produce more photons from dark matter annihilation, are also likely to produce more photons from astrophysical processes.  Furthermore, in an actual analysis on data, it may be possible to use multi-wavelength observations to infer the astrophysical (i.e.~non-dark matter) emission from a catalog of halos.  For example, for star forming galaxies, the star formation rate is expected to correlate with the gamma-ray luminosity; observations in the optical or infrared could therefore be used to place a prior on the gamma-ray luminosity of a halo \cite{Kornecki:2020}.   
Thus, while our assumed relation between astrophysical gamma-ray luminosity and halo mass is unlikely to be true in practice, it is reasonable to assume that we may have \textit{some} prior information about the gamma-ray luminosities of extragalactic halos.  Our analysis therefore approximates a future analysis of real data that includes such prior information.

We will thus assume that each halo contributes a flux 
of astrophysical photons given by 
\bea
\Phi_{\rm aniso} &=& C_{\rm aniso} 
\left(\frac{M_{\rm halo}}{M_\odot} \right) 
\left( \frac{D}{\mathrm{kpc}} \right)^{-2} ,
\eea
where $C_{\rm aniso}$ is a constant that we will fix below.

\subsection{Galactic Anisotropic Background}
\label{sec:gal_bg}

An important contribution to the gamma ray flux arises from cosmic ray interactions with the matter and radiation fields in our galaxy.
To estimate these backgrounds, we use the diffuse galactic emission model from Ref.~\cite{FermiCatalog4}, which fits the gamma ray flux observed by Fermi with a variety of source templates.\footnote{More details and images of the templates can be found at \url{https://fermi.gsfc.nasa.gov/ssc/data/analysis/software/aux/4fgl/Galactic_Diffuse_Emission_Model_for_the_4FGL_Catalog_Analysis.pdf}. We use the file  \texttt{gll\_iem\_v07.fits} for the galactic component.}

\subsection{Isotropic Background}
\label{sec:iso_background}

We expect that there will also be an additional gamma-ray
background which is roughly isotropic.  This 
background can be sourced by photon emission, either from 
astrophysical process or from dark matter annihilation, 
in the smooth component of the Milky Way halo (away from 
the Galactic Center and plane), or in unresolved extragalactic halos.  To this end, we use the isotropic background model produced by the Fermi team for front-converting events.\footnote{We use the file \texttt{iso\_P8R3\_SOURCE\_V3\_FRONT\_v1.txt}.}
This background estimate is derived in conjunction with the anisotropic background described above by fitting an isotropic template in the region outside of the galactic plane to avoid contamination.

Note that velocity-dependent dark matter annihilation in the smooth component of the Milky Way halo is not expected 
to be completely isotropic, even when looking away from the Galactic Center and plane (see, for example, Ref.~\cite{Lacroix:2022cjm}).  
In an actual analysis 
of data, one would include a template for either $p$-wave or $d$-wave dark matter annihilation within the Milky 
Way, with a normalization correlated with that of the extragalactic halo signal.  In this case, dark matter foreground 
emission may provide some additional statistical power, since the $p$-wave and $d$-wave templates will be different~\cite{Boddy:2018ike}. 
But in any case, we expect the effect to be relatively small, 
since the relative velocities of dark matter particles in the Milky Way are smaller than in extragalactic 
halos, and vary over a smaller range.  As we have noted, a complete treatment of backgrounds is beyond the scope 
of this work.

\section{Mock data analysis}
\label{sec:mock_data}

We begin by creating a mock data set, which consists of 
the number of photon counts in pixels across the survey footprint. 
The survey footprint (Fig.~\ref{fig:survey_geometry_skymap}) used for the mock analysis covers approximately 18\% of the sky, amounting to $N_{\rm pix} = 9012$ pixels in a \texttt{Healpix}\footnote{\url{https://healpix.jpl.nasa.gov/}} map with $N_{\rm side} = 64$, corresponding to roughly $1\,{\rm deg}^2$ pixels.  
We consider an instrument similar to the Fermi Large Area Telescope (Fermi-LAT), taking 
the exposure to be $10^4~\cm^2~\yr$, with $E_{\rm min} = 
1~\gev$ and $E_{\rm max} = 100~\gev$.  The resolution of Fermi-LAT corresponds roughly the our chosen pixel scale; at this resolution, the annihilation signals from the extragalatic halos are effectively point-like.

In the $i$th pixel, the expected number of photons with 
energies between $E_{\rm min}$ and $E_{\rm max}$ from 
dark matter annihilation is given by
\bea
N_{{\rm DM}(n);i} &=&  C_{{\rm DM}(n)} \sum_{h \in i}
\left(\frac{M_{s(h)}}{M_\odot} \right)^{2+(n/2)}
\left(\frac{r_{s(h)}}{\mpc}\right)^{-3-(n/2)}
\left(\frac{D_{(h)}}{\mpc} \right)^{-2} ,
\eea
where $M_{s(h)}$, $r_{s(h)}$ and $D_{(h)}$ are the scale mass, scale radius, 
and luminosity distance to the $h$th halo, respectively.  The sum is over all halos in pixel 
$i$, and 
\bea
C_{{\rm DM}(n)} &=&
1.8  
\left(6.1 \times 10^{-19}   \right)^{n/2} 
\times \tilde J_n^{\rm tot} 
\times \frac{\Phi_{PP}}{(\gev/ c^2)^{-2} \cm^3 \s^{-1} } 
\times \frac{{\rm exposure}}{\cm^2 \yr} .
\eea

For the case of $s$-wave dark matter annihilation 
($n=0$), observations of 
dwarf spheroidal galaxies (dSphs) require 
$\Phi_{PP(n=0)} \lesssim 
10^{-30}~(\gev /c^2)^{-2}~\cm^3~\s^{-1}$~\cite{Geringer-Sameth:2011wse,Boddy:2018qur,Boddy:2019kuw}.  
We also have $\tilde J_{n=0}^{\rm tot} = 1/3$. 
If we assume that $\Phi_{PP(n=0)}$ is at the upper bound allowed by 
searches of dSphs, then the expected number of photons 
arising from $s$-wave dark matter annihilation in all the extragalactic 
halos in our sample would then be given by 
$\sum_i N_{{\rm DM}(n=0);i} \sim {\cal O}(10^2)$.  It would be impractical to 
detect a signal involving such a small number of photons, given the expected 
backgrounds.  Essentially, we find that if $\Phi_{PP}$ is 
chosen so that the number of expected photons arriving from dSphs saturates the limit from current Fermi data, then 
the number of expected photons from extragalactic halos  
would be too small to be detected, assuming a boost 
factor of 1.  In order for $s$-wave annihilation to 
provide a detectable signal, one would need a large 
boost factor for extragalactic halos,
as assumed in Ref.~\cite{Regis:2015zka}.  Similar considerations would apply to the case of 
Sommerfeld-enhanced annihilation in the Coulomb 
limit ($n=-1$).  In that case, the number of 
photons arriving from extragalactic halos would be 
even more suppressed relative to those arriving from 
dSphs, in which the dark matter particles are moving 
more slowly.

Instead, we consider the case in which the true model is 
$p$-wave annihilation ($n=2$).  In this case, the 
constraint on $\Phi_{PP}$ from dSphs is about 8 orders of magnitude 
weaker~\cite{Boddy:2019qak}.  
If we take $\Phi_{PP(n=2)} = 10^{-22}~(\gev /c^2)^{-2}~\cm^3~\s^{-1}$, the expected total number of photons from dark matter annihilation in all halos in our catalog
is ${\cal O}(10^4)$\footnote{As an example, if we consider a model in which dark matter with 
$m_X = 10~\gev$ annihilates to $\bar b b$ from a $p$-wave state, this value of $\Phi_{PP}$ would 
correspond to $(\sigma v)_0 \sim 2.7 \times 10^{-19} \cm^3/s$.}.  For simplicity, we choose the 
overall normalization of the dark matter signal so that 
the expected number of photons from all pixels in our analysis due to dark matter annihilation in extragalactic halos is 5000.

Given our model for the extragalactic anisotropic background (\S\ref{sec:exgal_aniso}), the expected number of photons in the $i$th pixel due to anisotropic astrophysical processes is 
\bea
N_{{\rm aniso};i} &=&  C_{\rm aniso} \sum_{h \in i}
\left(\frac{M_{h}}{M_\odot} \right) 
\left( \frac{D_h}{\mathrm{kpc}} \right)^{-2} ,
\eea
where $C_{\rm aniso}$ is a normalization constant. The value of this constant will depend on details of the complex emission processes that give rise to astrophysical gamma-rays, as well as the extent to which multi-wavelength observations can be used to identify and remove regions of high astrophysical emission, such as blazars.  As a convenient benchmark for 
testing our ability to distinguish the dark matter 
signal from backgrounds, we choose $C_{\rm aniso}$
such that the expected number of photons produced by astrophysical process in extragalactic halos across all pixels is the same as for the dark matter signal, i.e.~5000 photons. 

The total number of photons expected in each pixel is 
thus 
\bea
N_{i} &=& N_{{\rm DM}(n);i} + N_{{\rm aniso};i} + N_{{\rm iso};i}
+ N_{{\rm gal};i},
\eea
where $N_{{\rm iso};i}$ and $N_{{\rm gal};i}$ are the expected number of 
photons in the $i$th pixel due to isotropic backgrounds (\S\ref{sec:iso_background}) and galactic backgrounds (\S\ref{sec:gal_bg}), respectively.
Using the isotropic background model discussed in \S\ref{sec:iso_background}, $N_{{\rm iso};i}$  is 
found to be $40.3$ for all pixels.  We assume that the photon count distribution 
in each pixel is Poisson, with the mean count given 
by the expected number of photons.  

For a real gamma-ray telescope, uncertainty in measured photon directions will lead to a smearing of the observed gamma-ray sky, changing the expectation values in each pixel.  For simplicity, we ignore this effect here.  Our pixel size of $1\,{\rm deg}^2$ is large enough that this is a sufficient approximation for Fermi-LAT over the energy range that we consider.  In any case, since such smearing amounts to a small change in the assumed templates, we do not expect it to have a significant impact on the results of our likelihood analysis.

\begin{figure*}
\begin{subfigure}[b]{0.45\textwidth}
\includegraphics[width=\linewidth]{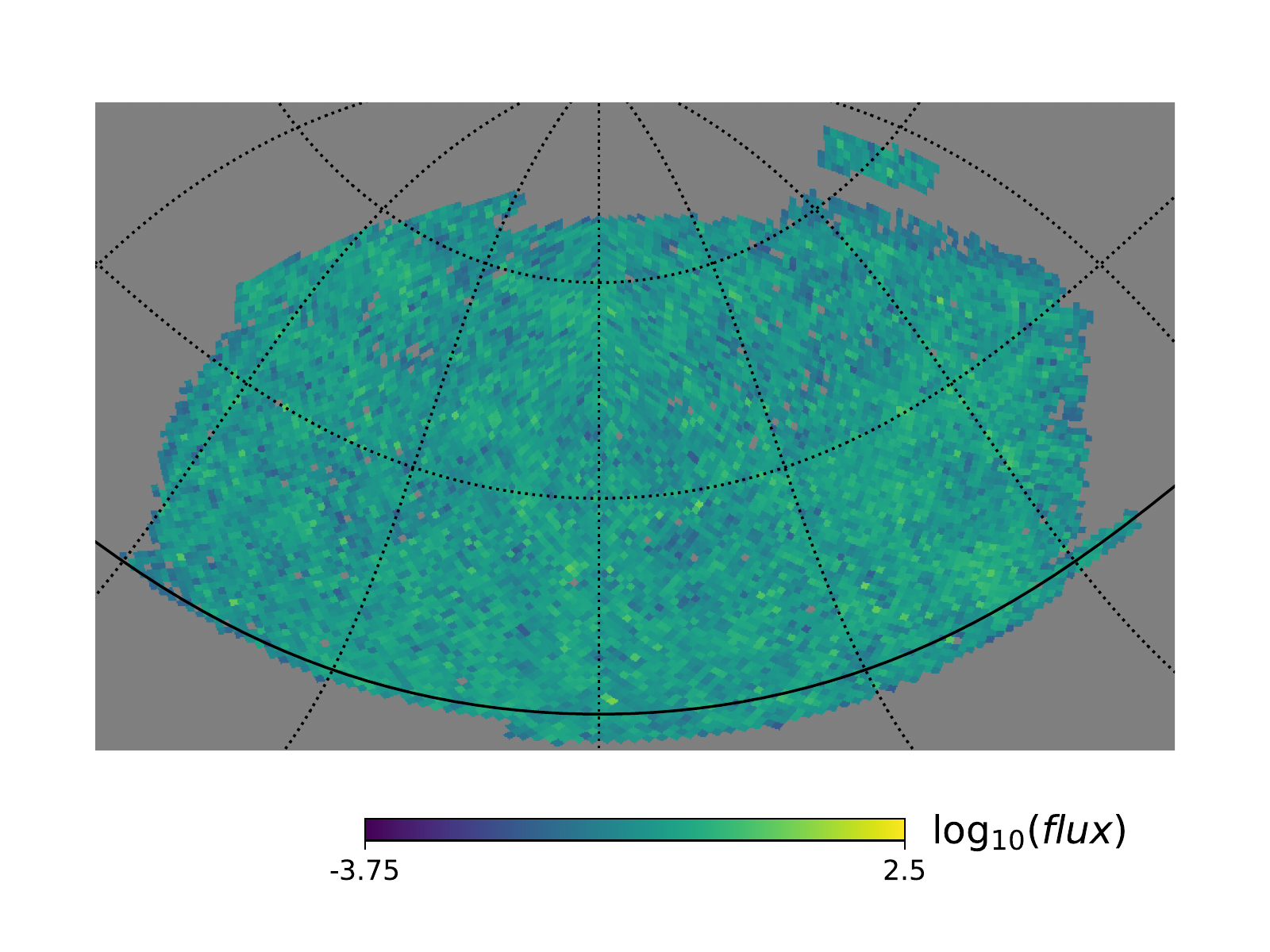}
\caption{$n=0$\label{fig:n=0signal_skymap}}
\end{subfigure}
\begin{subfigure}[b]{0.45\textwidth}
\includegraphics[width=\linewidth]{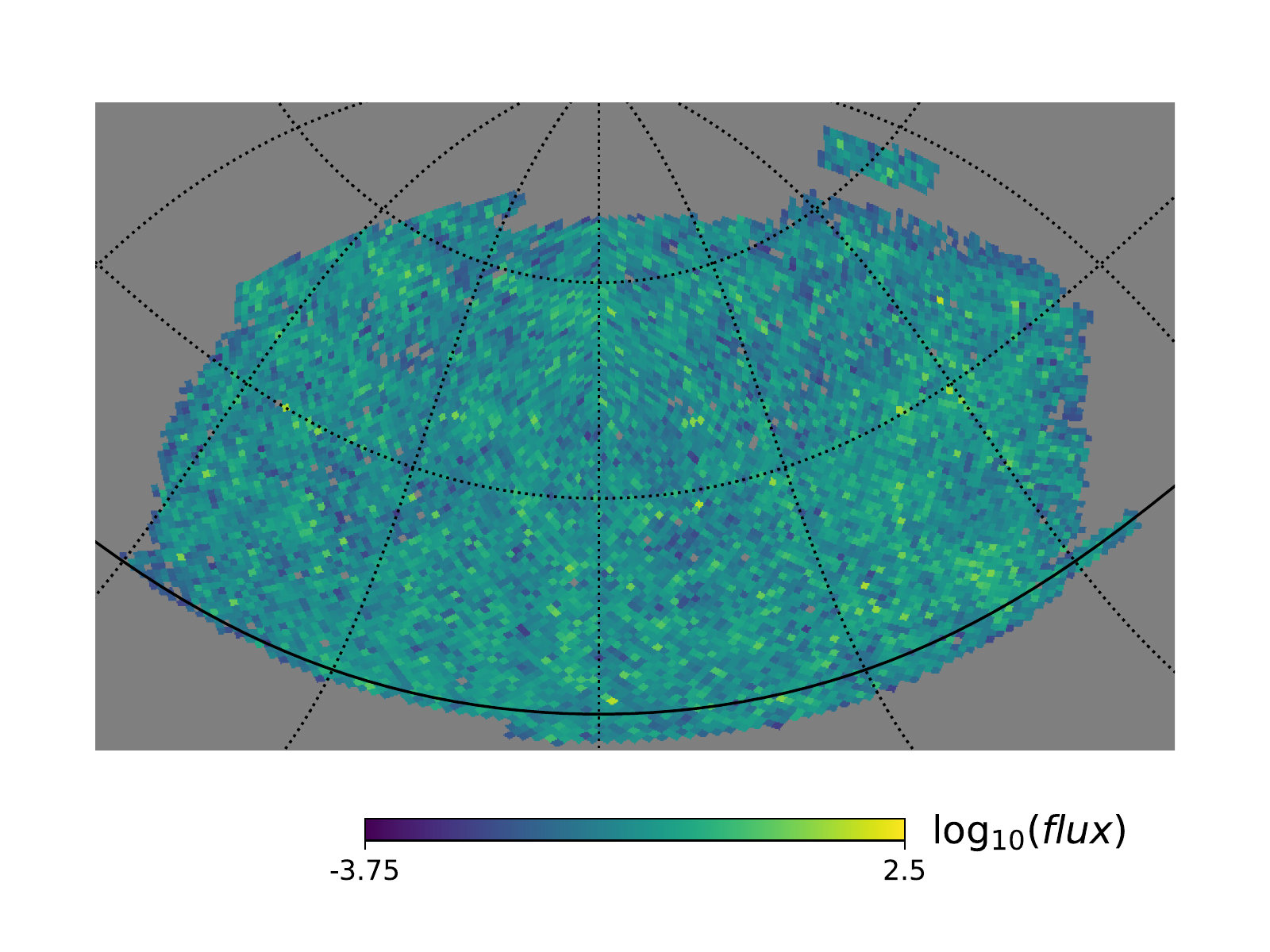}
\caption{$n=2$\label{fig:n=2signal_skymap}}
\end{subfigure}
\begin{subfigure}[b]{0.45\textwidth}
\includegraphics[width=\linewidth]{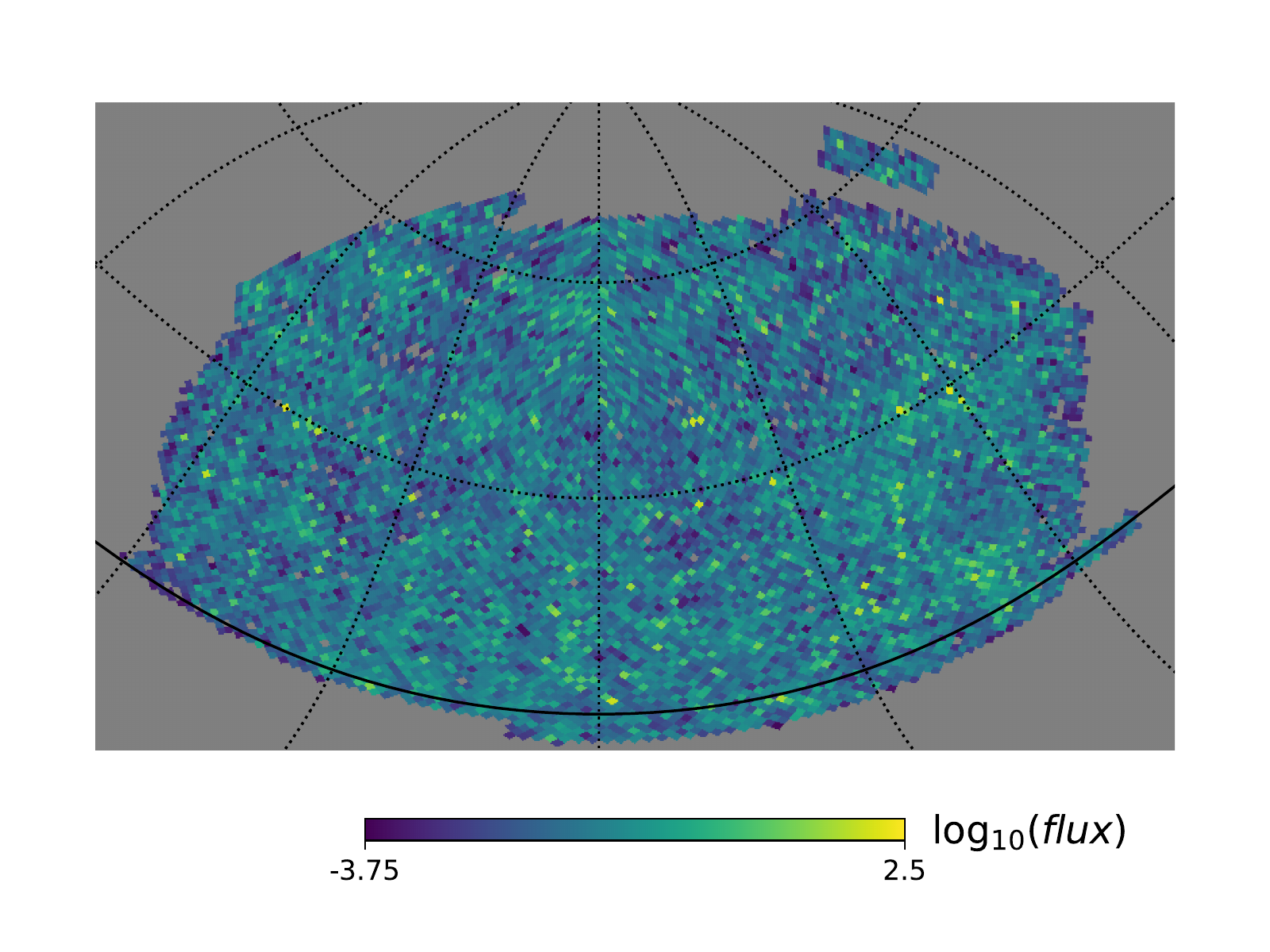}
\caption{$n=4$\label{fig:n=4signal_skymap}}
\end{subfigure}
\begin{subfigure}[b]{0.45\textwidth}
\includegraphics[width=\linewidth]{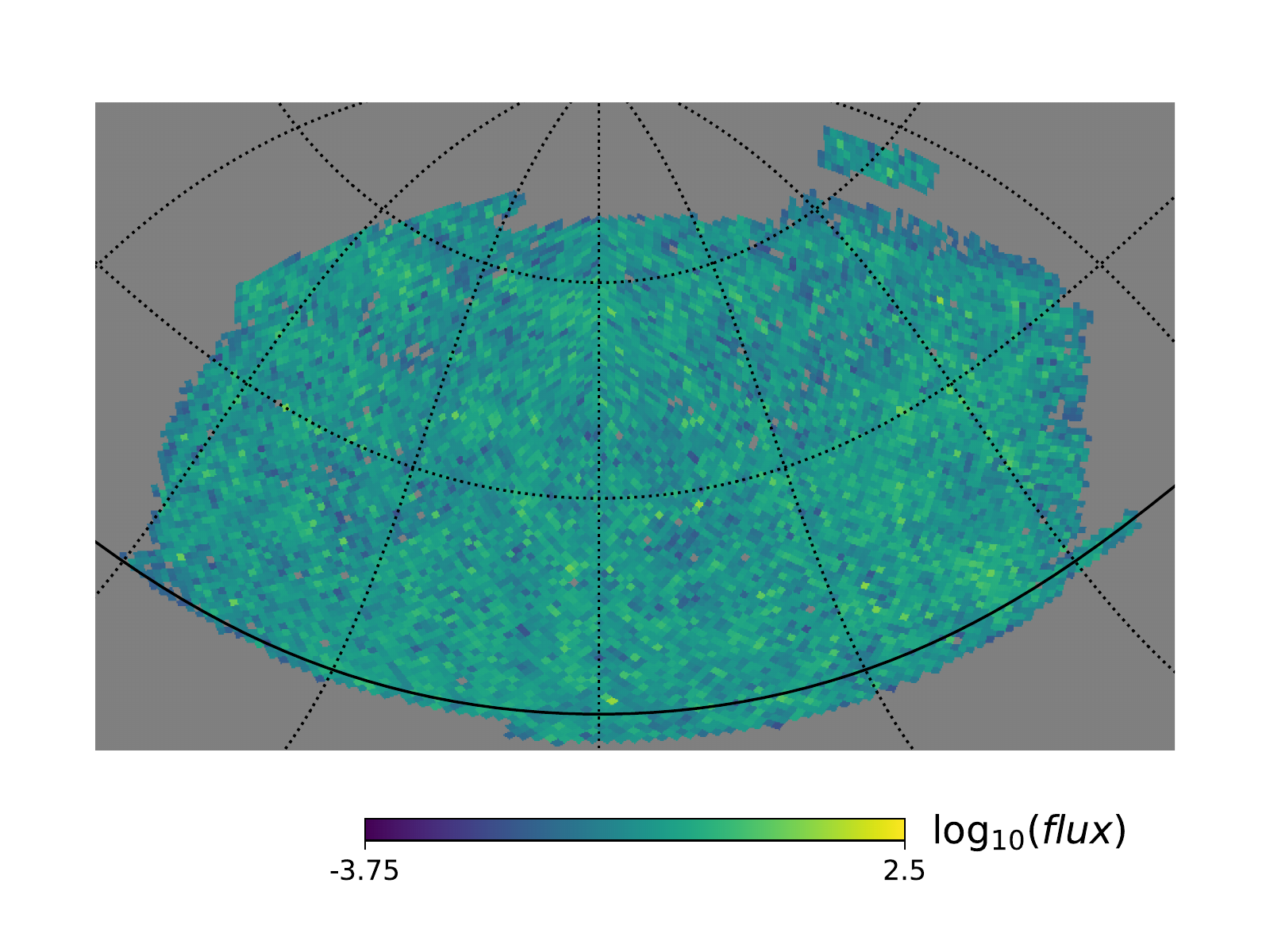}
\caption{anisotropic background.\label{fig:aniso_skymap}}
\end{subfigure}
\begin{subfigure}{\textwidth}
\centering
\includegraphics[width=0.45\linewidth]{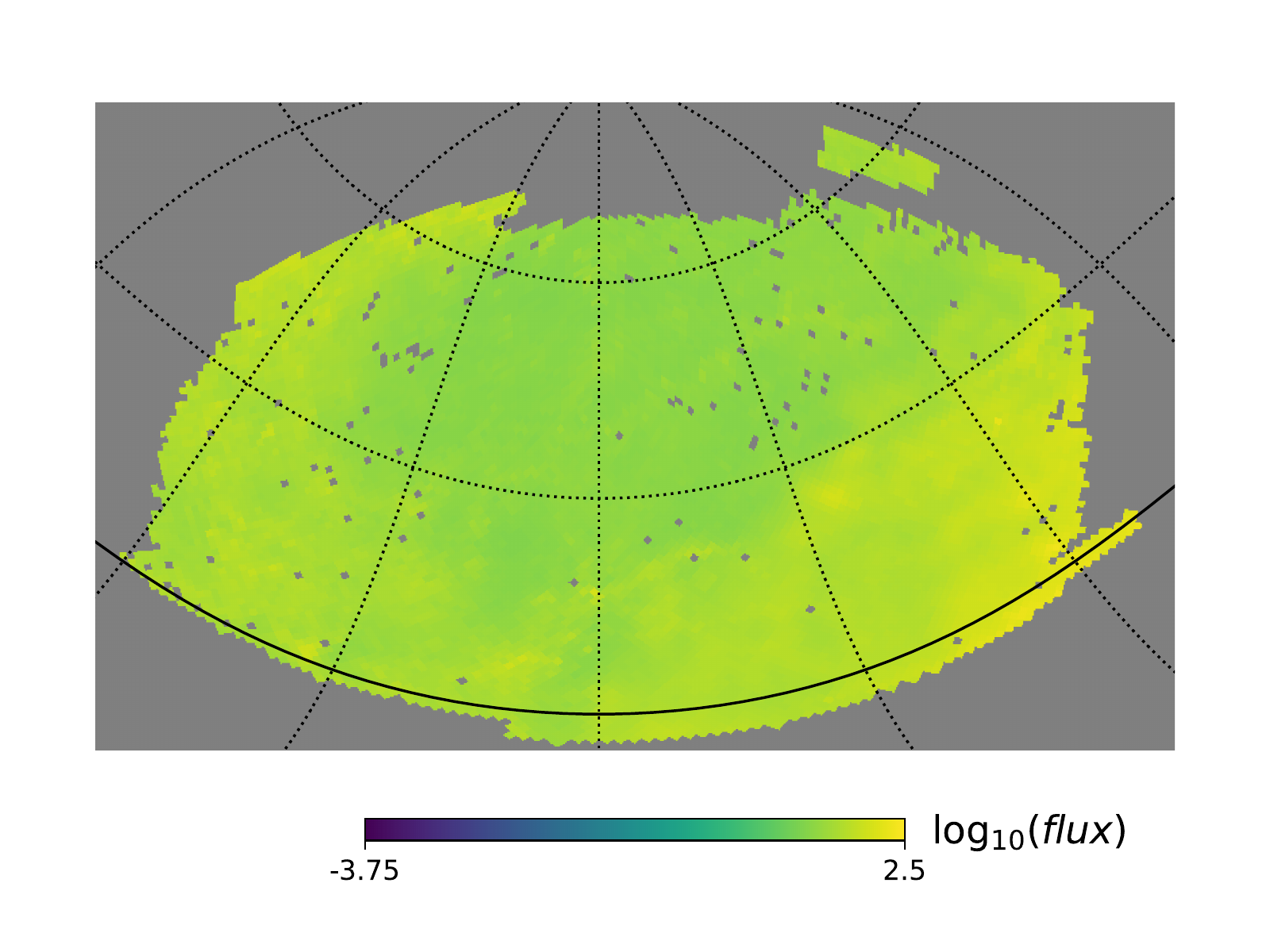}
\caption{Galactic background.\label{fig:fermi_skymap}}
\end{subfigure}
\caption{The flux map models used in our analysis.  Panels (a)--(c) represent the dark matter signal model with different values of the velocity-dependence parameter, $n$.  Panel (d) represents our toy model for the non-dark matter emission from extragalactic halos, while panel (e) represents the galactic background model.  Flux values are in units of photons per pixel, assuming our nominal exposure of $10^4~\cm^2~\yr$ and a pixel size corresponding to a \texttt{Healpix} resolution of $N_{\rm side} = 64$.} 
\end{figure*}

For illustration, we present the dark matter annihilation 
flux maps for $n=0$, 
$n=2$ and $n=4$ in Figures~\ref{fig:n=0signal_skymap},~\ref{fig:n=2signal_skymap}, 
and~\ref{fig:n=4signal_skymap}, respectively.  
In each case, 
we show signal only, and normalize the total 
expected number of photons from dark matter annihilation in 
extragalactic halos 
to $5000$, which, given an exposure of $10^4~\cm^2~\yr$, would be about the largest number of photons 
consistent with bounds from searches of dSphs, 
in the case of $p$-wave annihilation.  Note that for $n=0$, such a large photon count from 
extragalactic halos would be inconsistent with bounds from 
observations of dSphs unless the extragalactic halos had a 
sizeable boost factor.  
Similarly, we note that for $n=4$, models consistent with 
searches of dSphs could produce many more than $5000$ photons in 
extragalactic halos.  We plot the extragalactic anisotropic 
background flux map
in Figure~\ref{fig:aniso_skymap}, 
assuming the total expected number of photons is 5000, 
equal to the assumed signal for an exposure of 
$10^4~\cm^2~\yr$.  We plot 
the galactic anisotropic 
background flux map in the energy range $1-100~\gev$ in Figure~\ref{fig:fermi_skymap}, normalized to the expected number of counts per 
pixel for the same exposure.

With the mock data generated, we now compute the likelihood of data as a function of our model parameters.   We can completely describe the model for the photon probability 
distribution with the parameters $n$, $N_{{\rm DM}} = \sum_i N_{{\rm DM}(n);i}$, 
$N_{{\rm aniso}} = \sum_i N_{{\rm aniso};i}$ and $N_{\rm iso} = N_{{\rm pix}} N_{{\rm iso};i}$.
The likelihood for the observed data is given by a product of Poisson likelihood across all pixels:
\begin{equation}
\label{eq:likelihood}
    \mathcal{L}(\{d_0, d_1, \ldots, d_{N_{\rm pix}} \} | \vec{\theta}) = \prod_{i}^{N_{\rm pix}} \frac{N_{i}^{d_i} e^{-N_{i}}}{d_i!},
\end{equation}
where $d_i$ is the number of photons observed in the $i$th pixel and $\vec{\theta} = (n, N_{\rm DM}, N_{\rm aniso}, N_{\rm iso})$ is the set of model parameters.  The likelihood will provide a way to measure the signal amplitude, $N_{\rm DM}$, in the mock data, as well as a way to distinguish between different values of $n$.   In writing Eq.~\ref{eq:likelihood}, we have ignored the impact of the limited resolution of the telescope, which will induce a correlation between neighboring pixels; this approximation is justified for the reasons discussed above.  We assume uniform priors on all of the model parameters, so the posterior on these parameters is proportional to the likelihood.

\section{Results}
\label{sec:results}

\begin{figure}
\centering
  \includegraphics[width=0.6\textwidth]{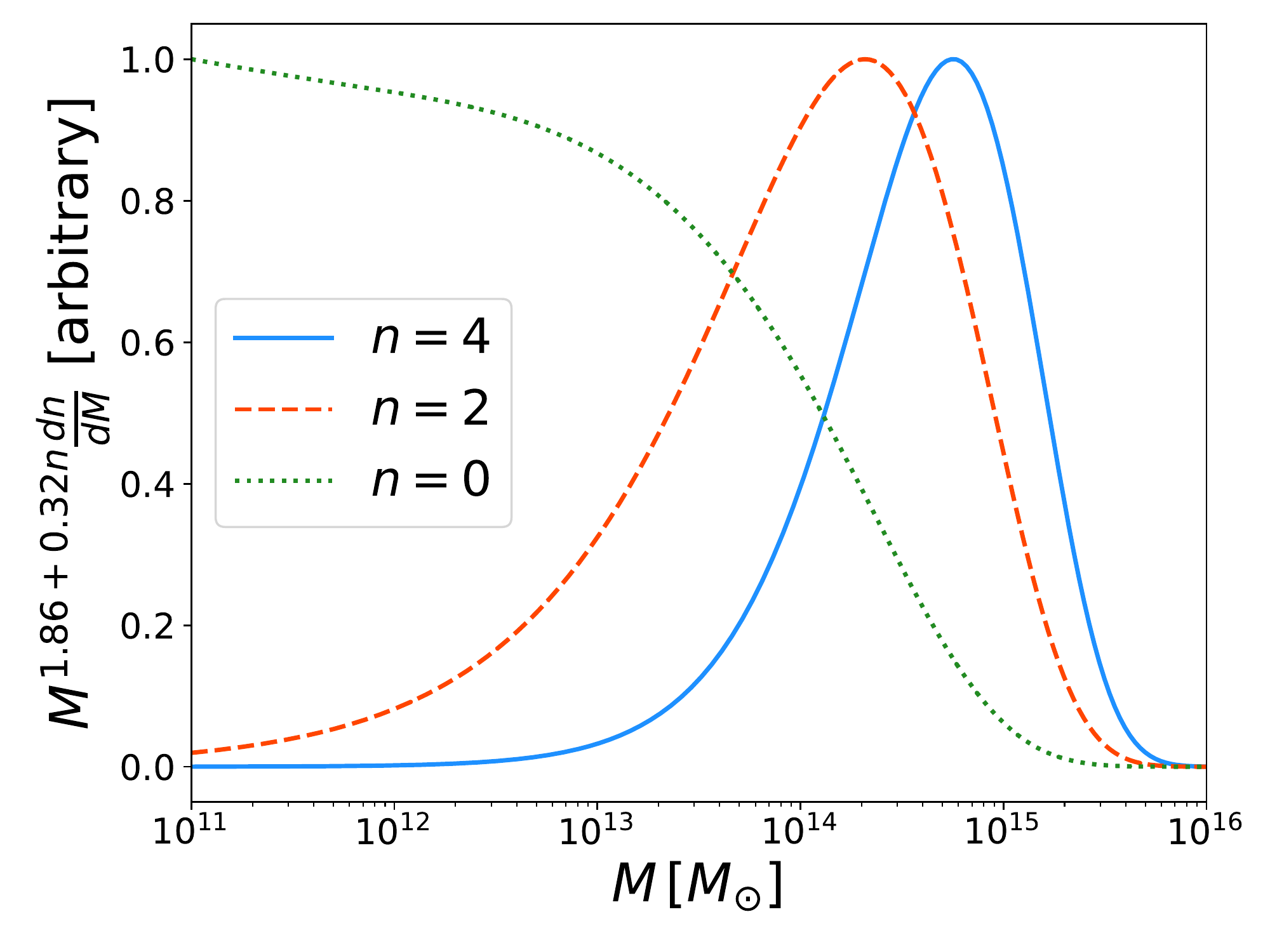}
\caption{Relative contributions to the annihilation luminosity as a function of halo mass for different models of the annihilation velocity dependence.  The annihilation luminosity is expected to scale as $M^{0.86 + 0.32n}$, where $M$ is the halo mass.  The $y$-axis thus represents the contribution to the luminosity per logarithmic interval of mass, normalized to unity.  The calculation is performed at $z = 0$ and assumes the halo mass function ($dn/dM$) from Ref.~\cite{Tinker:2008}.}
    \label{fig:mass_dependence}
\end{figure}

\subsection{Halo mass dependence}
\label{sec:halo_mass_dependence}

Before analyzing the mock data described above, we first compute the relative contributions of different halo masses to the expected extragalactic annihilation signal.  We assume a flat $\Lambda$CDM cosmological model with parameters matching those preferred by {\it Planck} \cite{Planck18}.  To describe the abundance of dark matter halos as a function of mass and redshift, we adopt the halo mass function from Ref.~\cite{Tinker:2008}.  Using our expression Eq.~\ref{eq:simplified_J} for the $J$-factor, we plot the relative contributions to the annihilation signal from logarithmic halo mass bins at $z=0$ in Figure~\ref{fig:mass_dependence}.  

Consistent with the expectations discussed previously, as $n$ is increased, the halo mass range contributing to the total annihilation luminosity shifts towards higher masses.  For the models that we focus on in this analysis ($n=2$ and $n=4$), most of the annihilation luminosity comes from halos above the group scale ($M \sim 10^{13} M_{\odot}$).  As noted previously, the catalog described in \S\ref{sec:halocatalog} therefore captures most of the annihilation signal relevant for the models of interest.

\subsection{Mock analysis}

Having drawn mock data assuming a true model of 
$p$-wave annihilation ($n=2$) and an exposure 
of $10^4~\cm^2~\yr$,
we compute the likelihood 
of the data across a grid of parameter values, assuming either the $n=2$ or the $n=4$ model; we will use these grids to plot marginalized posteriors.  Additionally, we use the Nelder-Mead algorithm
to maximize the likelihood and determine the best-fit model parameters.

\begin{figure*}
\centering
\begin{subfigure}[b]{0.75\textwidth}
\includegraphics[width=\linewidth]{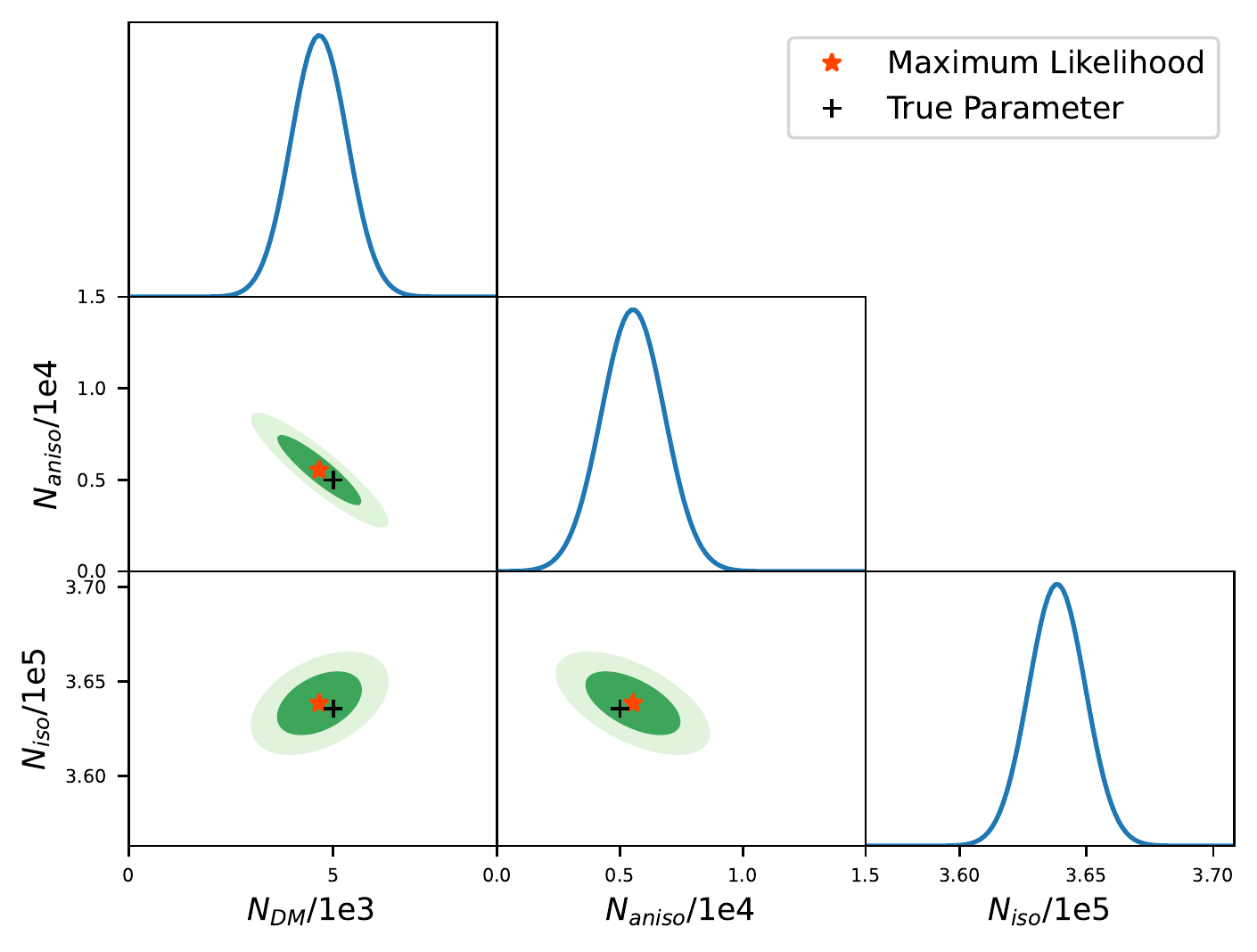}
\caption{Data: $n=2$, Model: $n=2$.\label{fig:n=2model_n=2data_noboost}}
\end{subfigure}
\vfill
\begin{subfigure}[b]{0.75\textwidth}
\includegraphics[width=\linewidth]{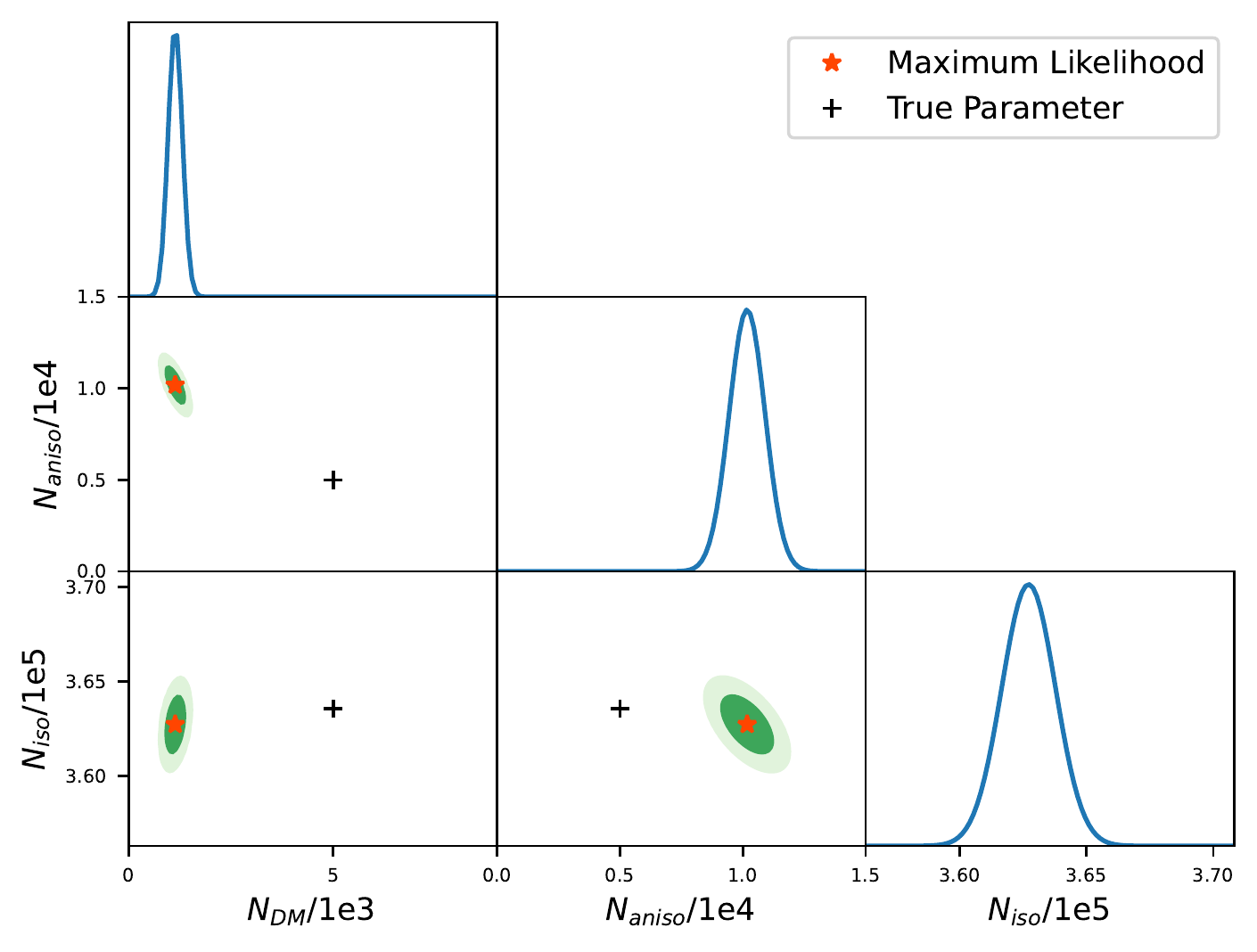}
\caption{Data: $n=2$, Model: $n=4$.\label{fig:n=4model_n=2data_noboost}}
\end{subfigure}
\caption{Results of the likelihood analysis of mock gamma-ray data.  The top panel (a) shows the results of analyzing mock data generated assuming $n=2$ with the correct velocity-dependence model; bottom panel (b) shows the results of analyzing the same mock data with a different, $n=4$, velocity-dependence model.  We assume an exposure of 
$10^4~\cm^2~\yr$, roughly equal to the current exposure of Fermi.  Contours indicate the 68\% (dark green) and 95\% (light green) credible intervals on the model parameters. Red cross hairs represent the location of the maximum likelihood (see Table \ref{tab:results_table_noboost}). Black cross hairs indicate the normalization of the true model: [$N_{\rm DM}$, $N_{\rm aniso}$, $N_{\rm iso}$] = [5000, 5000, $363550$]. \label{fig:likelihood_noboost} }
\end{figure*}

We  plot marginalized parameter constraints, assuming the 
correct model ($n=2$), 
in Figure~\ref{fig:n=2model_n=2data_noboost}, and assuming 
the incorrect model ($n=4$) in Figure~\ref{fig:n=4model_n=2data_noboost}.  In both figures, 
the true model is denoted with black crosses and the 
model which maximizes the likelihood is denoted with 
red crosses, and we take the exposure to be 
$10^4~\cm^2~\yr$, corresponding roughly to the current exposure with Fermi.  As discussed above, we have assumed that astrophysical backgrounds associated with extragalactic halos have been cleaned to a level comparable to the dark matter signal.  Making these assumptions, it is clear that with the current Fermi 
exposure, $N_{\rm DM} = 0$ is excluded at high significance, meaning that one can easily reject the null hypothesis of no 
dark matter contribution ($\Delta \ln \mathcal{L} = 24.0$).  As expected, we do well at reconstructing 
the parameters when we assume the correct model.  
In particular, this indicates that the 
$\sim 10^4$  photons one would observe contain enough 
information to provide very strong evidence for 
dark matter annihilation, even allowing for a floating 
background which is correlated with halo mass and an isotropic background of unknown amplitude.  
But if we assume an incorrect velocity-dependence for the 
dark matter annihilation cross section, the best-fit 
point is quite far from the true model, though interestingly, an astrophysics-only explanation (i.e. $N_{\rm DM} = 0$) would 
still be somewhat disfavored.

The difference between the maximum log-likelihoods for the $n=2$ and $n=4$ analyses, $\Delta \ln \mathcal{L}$, and corresponding best-fit parameter values are reported in Table~\ref{tab:results_table_noboost}.  For an exposure comparable to Fermi, we find that the difference in maximum likelihoods is too small to distinguish between the $n=2$ and $n=4$ models at high significance ($\Delta \ln \mathcal{L} = 1.1$).

\begin{table}[H]
    \centering
    \begin{tabular}{| l | r | r | r |} \hline
        Exposure $=10^{4}\,{\rm cm}^2~{\rm yr}$ & $n=2$ model & $n=4$ model & No DM \\ \hline \hline
        $\Delta \ln \mathcal{L}$ & 0 & 1.1 & 24.0 \\ \hline
        $N_{\rm DM}$ at maximum likelihood & 4654.6 & 1138.1 & 0 \\ \hline
        $N_{\rm aniso}$ at maximum likelihood & 5536.8 & 10179.8 & 13578.7 \\ \hline
        $N_{\rm iso}$ at maximum likelihood & 363835.6 & 362714.0 & 360469.3 \\ \hline
    \end{tabular}
    \caption{Results of likelihood analyses when the true model is $n=2$.  
    The reported $\Delta \ln \mathcal{L}$ values are measured with respect to the $n=2$ analysis.
    We assume an exposure of $10^4~\cm^2~\yr$ and the true parameter values are $N_{\rm DM} = 5000$, $N_{\rm aniso} = 5000$, and $N_{\rm iso} = 363550$. 
    }
    \label{tab:results_table_noboost}
\end{table}

\begin{figure*}
\centering
\begin{subfigure}[b]{0.75\textwidth}
\includegraphics[width=\linewidth]{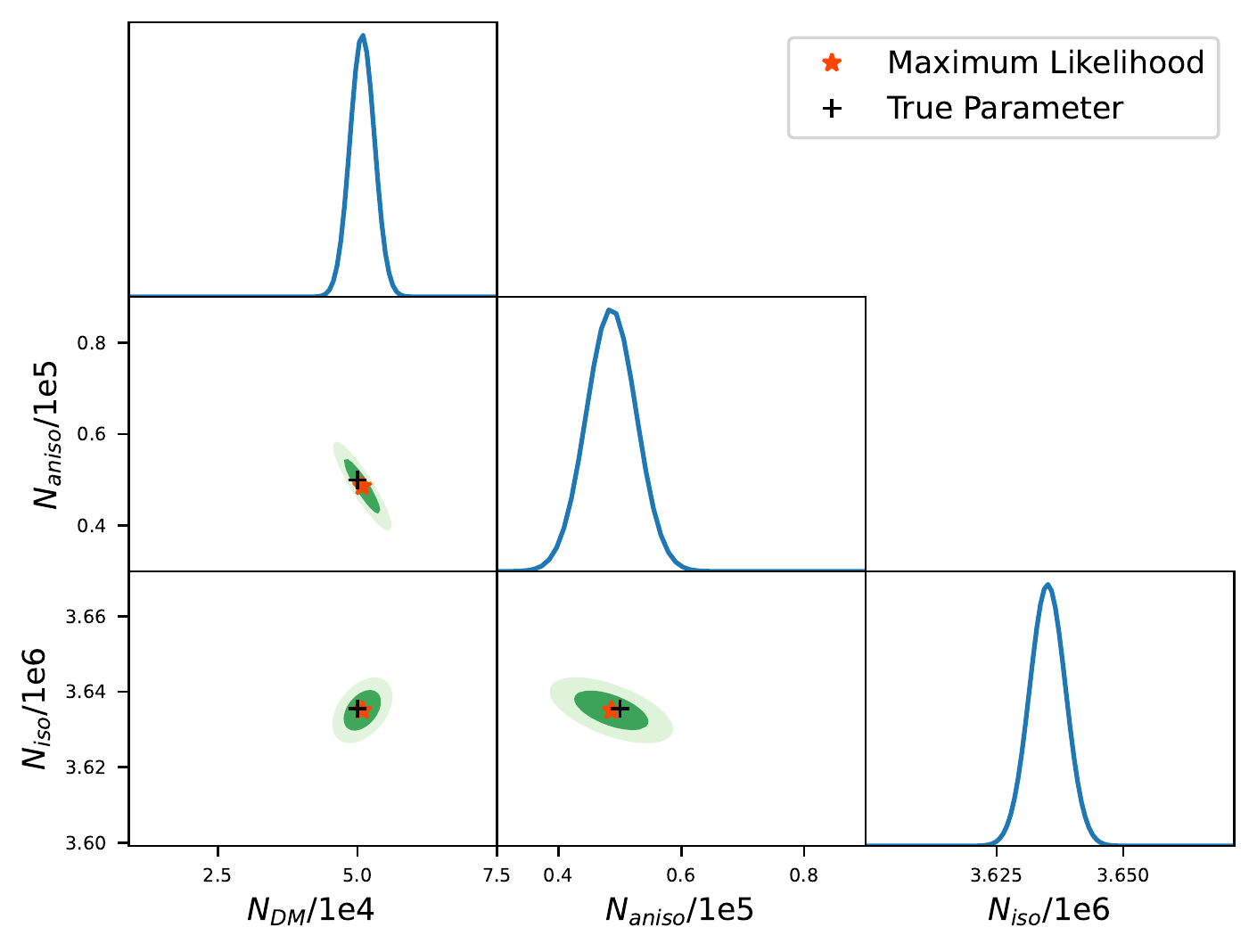}
\caption{Data: $n=2$, Model: $n=2$.\label{fig:n=2model_n=2data_boost}}
\end{subfigure}
\vfill
\begin{subfigure}[b]{0.75\textwidth}
\includegraphics[width=\linewidth]{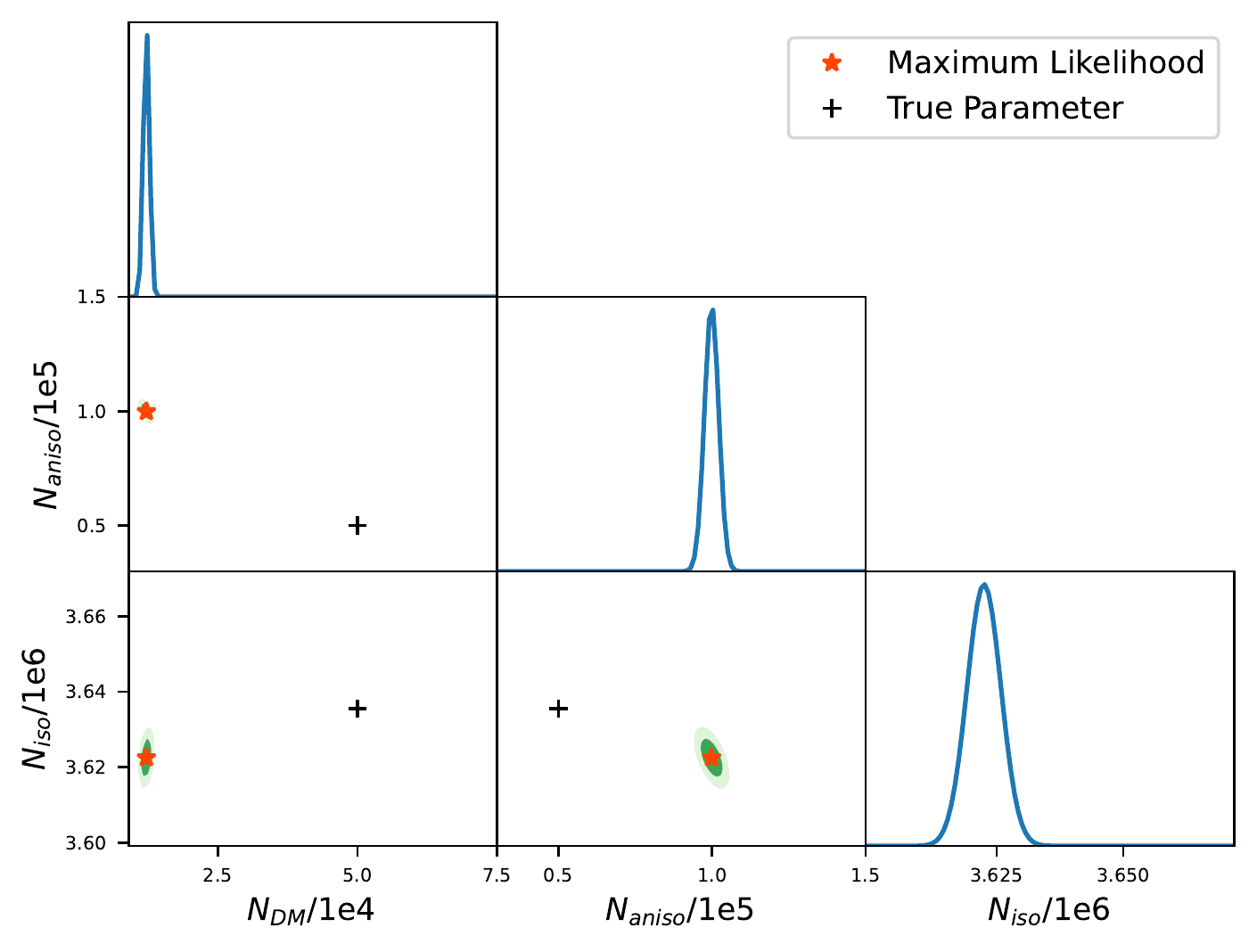}
\caption{Data: $n=2$, Model: $n=4$\label{fig:n=4model_n=2data_boost}}
\end{subfigure}
\caption{Same as Fig.~\ref{fig:likelihood_noboost}, but for an exposure of $10^5~\cm^2~\yr$, which is roughly $10\times$ larger than currently achieved by Fermi.  Red cross hairs represent the location of the maximum likelihood (see Table \ref{tab:results_table_boost}). Black cross hairs indicate the normalization of the true model: [$N_{\rm DM}$, $N_{\rm aniso}$, $N_{\rm iso}$] = [50000, 50000, $3635497$].  In some cases, the contours are so small that they are not visible below the markers.}
\end{figure*}

We then perform a similar analysis, drawing mock data 
assuming a true model with $n=2$, but with an exposure 
$10 \times$ larger ($10^5~\cm^2~\yr$).  
We present the resulting $\Delta \ln \mathcal{L}$ and best-fit
normalization parameters in Table~\ref{tab:results_table_boost}.
With a factor 10 larger exposure, one can clearly distinguish the true velocity-dependence model: $\Delta \ln \mathcal{L} = 22.5$.  We plot parameter constraints, assuming the 
correct model ($n=2$), 
in Figure~\ref{fig:n=2model_n=2data_boost}, and assuming 
the incorrect model ($n=4$) in Figure~\ref{fig:n=4model_n=2data_boost}.  In both figures, 
the true model is denoted with black crosses and the 
model which maximizes the likelihood is denoted with 
red crosses, and we take the exposure to be 
$10^5~\cm^2~\yr$.   Again, the true parameters are recovered to within the uncertainties when analyzing the data with the correct model.  

Obtaining an exposure roughly ten times larger than the current {\it Fermi} LAT exposure with the same instrument is unlikely.  Therefore, it seems that a new telescope (likely a larger instrument, or multiple instruments) will be needed to obtain the photon statistics necessary to distinguish between the $n=2$ and $n=4$ models in the manner we have considered above.  To investigate the required properties of such a telescope, we repeat our analysis across a grid of exposure values to build an interpolating function between exposure and $\Delta \ln \mathcal{L}$.  A reasonable threshold for rejecting one value of $n$ over another is $\Delta \ln \mathcal{L} \sim 10$.  For instance, if the models are compared using the Akaike information criterion \cite{Elements}, this would mean that one model is $\exp (10/2) \approx 150$ times more likely to minimize the information lost by using that model to represent the data.  We find that achieving  $\Delta \ln \mathcal{L} = 10$ would require an exposure 5.4 times larger than {\it Fermi} LAT.

\begin{table}[H]
    \centering
    \begin{tabular}{| l | r | r | r |} \hline
        Exposure $=10^{5}\,{\rm cm}^2~{\rm yr}$  & $n=2$ model & $n=4$ model & No DM \\ \hline \hline
        $\Delta \ln \mathcal{L}$ & 0 & 22.5 & 284.3 \\ \hline
        $N_{\rm DM}$ at maximum likelihood & 50845.9 & 12189.4 & 0 \\ \hline
        $N_{\rm aniso}$ at maximum likelihood  & 48606.7 & 99868.2 & 136061.5\\ \hline
        $N_{\rm iso}$ at maximum likelihood & 3635062.4 & 3622512.9 & 3598680.5 \\ \hline
    \end{tabular}
    \caption{Same as Table~\ref{tab:results_table_noboost}, but with $10\times$ larger exposure.  The larger exposure means the true parameter values are now  $N_{\rm DM} = 50000$, $N_{\rm aniso} = 50000$, and $N_{\rm iso} = 3635497$.
    }
    \label{tab:results_table_boost}
\end{table}

\section{Conclusion}
\label{sec:conclusion}

We have considered the indirect detection of velocity-dependent dark matter annihilation in extragalactic halos using gamma-ray signals.  Extragalactic halos are an especially interesting 
target because, of all bound astrophysical objects, 
they have among the largest particle velocities.  
Thus, if  $\sigma v \propto (v/c)^n$ for $n>0$, one 
expects the dark matter annihilation signal from 
extragalactic halos to be enhanced relative to other 
commonly-studied targets, such as galactic subhalos, 
dwarf spheroidal galaxies, and the Galactic Center.

Our analysis makes several optimistic assumptions, including that backgrounds are modelled perfectly.  In particular, 
although we incorporate anisotropic 
astrophysical backgrounds which are correlated with 
the halo map (arising from astrophysical processes 
in galaxies), we assume that the dependence of the 
astrophysical background flux on the halo mass is 
a simple proportionality, which can thus be 
distinguished from the dark matter annihilation signal.  

Even with these optimistic assumptions, we find 
that there is not enough information  in current Fermi data to detect photons produced by 
$s$-wave dark matter annihilation in the extragalactic halos in our catalog, assuming $\Phi_{PP}$ is set by current limits from dSphs.   Our analysis assumes no boost to the dark matter signal arising from the presence of subhalos within the main halos.  In the absence of a large boost factor, any model which could be observed in searches of extragalactic halos would already be ruled out by searches of dSphs.  

For the case of $p$-wave annihilation, although 
the current 
Fermi data contains enough information to reject the 
null hypothesis of no dark matter contribution, it 
only marginally 
favors the true model of annihilation over an allowed 
alternative model, such as $d$-wave annihilation.  But 
a factor of 10 larger exposure would allow one to 
distinguish the correct model of dark matter annihilation 
from the incorrect one with high significance.

Perhaps the most important avenue for future work is 
in the characterization of astrophysical backgrounds, 
particular those which could be correlated with the 
dark matter signal.  One might hope that future 
developments in multi-wavelength astronomy would allow 
one to correlate a gamma-ray signal from astrophysical 
processes in a halo with emission in other bands, 
resulting in a more accurate extragalactic anisotropic 
background map.  

{\bf Acknowledgements}

JK is supported in part by DOE grant DE-SC0010504.
JR is supported by NSF grant AST-1934744.  We thank an anonymous referee for comments on our manuscript.

\bibliography{thebibliography.bib}
\end{document}